\input harvmac
\input epsf

\def\p{\partial}
\def\ap{\alpha'}
\def\half{{1\over 2}}

\def\({\left(}
\def\){\right)}
\def\[{\left[}
\def\]{\right]}

\Title{}{\vbox{\centerline{Weak Gravity Conjecture }
\smallskip
\centerline{for}
\smallskip
\centerline {Noncommutative Field Theory}}}

\centerline{Qing-Guo Huang $^1$ and Jian-Huang She $^2$}

\bigskip \centerline{\it $^1$ School of physics, Korea Institute for
Advanced Study, } \centerline{\it 207-43, Cheongryangri-Dong,
Dongdaemun-Gu, Seoul 130-722, Korea}
\medskip
\centerline{\it $^2$ Institute of Theoretical Physics, Academia
Sinica} \centerline{\it P. O. Box 2735, Beijing 100080, P. R. China}

\bigskip
\centerline{\tt huangqg@kias.re.kr} \centerline{\tt jhshe@itp.ac.cn}
\bigskip

We investigate the weak gravity bounds on the U(1) gauge theory and
scalar field theories in various dimensional noncommutative space.
Many results are obtained, such as the upper bound on the
noncommutative scale $g_{YM}M_p$ for four dimensional noncommutative
U(1) gauge theory. We also discuss the weak gravity bounds on their
commutative counterparts. For example, our result on 4 dimensional
noncommutative U(1) gauge theory reduces in certain limit to its
commutative counterpart suggested by Arkani-Hamed et.al at least at
tree-level.

\Date{August, 2006}


\nref\vafa{C. Vafa, hep-th/0509212.}

\nref\amnv{N. Arkani-Hamed, L. Motl, A. Nicolis and C. Vafa,
hep-th/0601001. }

\nref\kms{S. Kachru, J. McGreevy and P. Svrcek, JHEP 0604(2006)023,
hep-th/0601111.}

\nref\lsw{M. Li, W. Song, T. Wang, hep-th/0601137. }

\nref\ov{H. Ooguri and C. Vafa, hep-th/0605264. }

\nref\hls{Q.G. Huang, M. Li, W. Song, hep-th/0603127; Q.G. Huang,
hep-th/0610106.}

\nref\kmp{Y. Kats, L. Motl, M. Padi, hep-th/0606100. }

\nref\lssw{M. Li, W. Song, Y.S. Song and T. Wang, hep-th/0606011; A.
Medved, hep-th/0611196.}

\nref\dl{M.R. Douglas, Z. Lu, hep-th/0509224; B.S. Acharya, M.R.
Douglas, hep-th/0606212. }

\nref\bjs{T. Banks, M. Johnson and A. Shomer, hep-th/0606277. }

\nref\sw{N. Seiberg, E. Witten, JHEP 9909 (1999) 032,
hep-th/9908142.}

\nref\gn{D.J. Gross and N.A. Nekrasov, JHEP 0007(2000)034,
hep-th/0005204; D.J. Gross and N.A. Nekrasov, JHEP 0010 (2000) 021,
hep-th/0007204. }

\nref\mst{A. Matusis, L. Susskind and N. Toumbas, JHEP
0012(2000)002, hep-th/0002075. }

\nref\dn{M.R. Douglas, N.A. Nekrasov, Rev.Mod.Phys. 73(2001)977,
hep-th/0106048.}

\nref\dk{G.M. Derrick, J.Math.Phys. 5(1964)1252. }

\nref\gms{R. Gopakumar, S. Minwalla and A. Strominger, JHEP
0005(2000)020, hep-th/0003160. }

\nref\jh{J.A. Harvey, hep-th/0102076. }

\nref\mrs{S. Minwalla, M.V. Raamsdonk and N. Seiberg, JHEP
0002(2000)020, hep-th/9912072. }

By now, low energy effective field theory has become the cornerstone
in our description of nature. One of the embarrassing features of
effective field theory is that couplings of the fields are
determined phenomenologically, beyond a first principal derivation.
It is generally expected that a quantum theory of gravity would shed
some light on this issue. Recently in searching for criteria to
distinguish string landscape, which has a well-defined UV theory,
from the swampland, which cannot be completed to a fully
self-consistent theory, it was proposed in \refs{\vafa, \amnv} that
gravity induces some constraints on the quantum field theory.

Field theory is generally well defined in the weak coupling limit.
However the situation changes when we take gravity into account,
since in the weak coupling limit the nonperturbative objects in the
field theory are very heavy and the gravitational effects on them
are significant. In \amnv, requiring that the magnetic monopole
should not collapse to be a black hole yields a new nontrivial UV
cutoff for the U(1) gauge theory. Some related works are discussed
in \refs{\kms-\bjs}.

In this note we generalize the results of \amnv\ to noncommutative
gauge theory and scalar field theory. In particular, the scalar
field theory in three or higher dimensions, which has no solitonic
solutions in commutative space, will possess such solutions when
promoted to noncommutative space. Thus the mechanism proposed in
\amnv\ can be more generically used for noncommutative field
theories. The investigation of noncommutative field theory can also
give us some hints on conventional field theory.

First we consider in this note type IIB string theory with a
D3-brane placed in a non-zero B field with components along the
space-space directions. It is convenient to define the open string
metric $G_{ij}$, constant asymmetric matrix $\theta^{ij}$, gauge
coupling $g_{YM}$ through closed string metric $g_{ij}$, B-field
$B_{ij}$ and string coupling $g_s$ as follows:
\eqn\os{\eqalign{G_{ij}&=g_{ij}-(2\pi\ap)^2(Bg^{-1}B)_{ij} \cr
\theta^{ij} &=-(2\pi\ap)^2({1\over{g+2\pi\ap B}}B{1\over{g-2\pi\ap
B}})^{ij} \cr g^2_{YM} &=2\pi g_s(\det(1+2\pi\ap
g^{-1}B))^{1\over2}.}} We choose the constant B-field to be $\half
Bdx^1\wedge dx^2$, with the D3-brane lying in 0123 directions and
set the open string metric to be Euclidean $G_{ij}=\delta_{ij}$,
thus we have \eqn\bbb{B={\theta\over{(2\pi\ap)^2+\theta^2}}}and
\eqn\ggg{g_s=g^2_{YM}{\ap\over{\sqrt{(2\pi\ap)^2+\theta^2}}}.}
According to \sw, in the limit $\ap\rightarrow0$ with $G, \theta,
g^2_{YM}$ fixed, closed string degrees of freedom as well as massive
open string degrees of freedom decouple, and the brane reduces to a
field theory, namely ${\cal N}=4$ super Yang-Mills theory on a
noncommutative space with \eqn\ta{[x^i,x^j]=i\theta^{ij}.} Here we
need to stress that we only pay attention to the constraints on the
field theory in the noncommutative space and don't need to take care
of the origin of the noncommutative space.

It is also tempting to go in the reversed direction, that is, to
start with the decoupled field theory, and turn on gravitational
effects gradually, and ask about the constraints imposed by gravity
upon the parameters in the field theory \amnv. In \amnv\ it was
shown that for a U(1) gauge theory in commutative space, unless the
cutoff satisfies \eqn\weak{\Lambda\leq g_{YM}M_4,} where $M_4\sim
{G_4}^{-1/2}$ is the Planck scale in four dimensions, the magnetic
monopole would collapse to a black hole, ruining the validity of the
effective field theory. The advantage of the methods used in \amnv\
is that the constraint does not depend on the particular form of the
high energy completion of the field theory, of which one still lacks
of full control. Subsequent work in this direction includes
\refs{\lsw, \hls}.

The U(1) NCYM also has a monopole solution, but with a semi-infinite
string attached to it \gn. In fact, the solution is everywhere
non-singular and the energy density localizes along a half-line. The
tension of the string is given by
\eqn\ten{T={{2\pi}\over{g^2_{YM}\theta}}.} This soliton is
qualitatively different from their commutative counterpart, which is
point-like, with no strings detached.

When one turns on gravity, a natural question to ask is how do
noncommutative field theories couple to gravity, or should they
couple to conventional gravity or some noncommutative version. In
our paper, we only take care of the IR behavior of gravity and thus
we assume the noncommutative field theories are coupled to the
conventional gravity.

Another question is whether the field theory solitons persist their
existence when quantum gravity effects are turned on. For general
cases we still do not have a general proof of their existence. But
for the case at hand, which has a natural embedding in string
theory, this question has a simple positive answer. The soliton
solution, a monopole with a string detached, is just a D1-string
ending on a D3-brane. And its tension can also be calculated from
the brane configuration to be exactly \ten\ in \gn.

From the point of view of an asymptotic observer, the effect of the
string is to produce a deficit angle $8\pi G_4 T$ in spacetime.
Requiring that the deficit angle is less than $2\pi$ yields
\eqn\defic{8\pi G_4 T \leq 2\pi.} Ignoring numerical factors, one
gets \eqn\aaa{g_{YM}^2\theta \geq G_4 \sim M_4^{-2}.} When
space-space becomes noncommutative the weak gravity conjecture
yields a lower bound on the gauge coupling; or equivalently, upper
bound on the noncommutative scale $M_{nc}=\theta^{-1/2}$ with
\eqn\unc{M_{nc}\leq g_{YM}M_4.} Naively to go to the commutative
theory, one chooses the cutoff scale $\Lambda$ to be lower than the
noncommutative scale $\Lambda \leq M_{nc}$, which leads to \weak.
However, this is reliable only in the tree-level approximation,
because of the non-analytic behavior of $\theta$ \mst\ in
noncommutative gauge theory. In supersymmetric gauge theories the
logarithmic divergences at small values of noncommutative momenta
typically appear. One can only expect that in ${\cal N}=4$
supersymmetric gauge theory even the logarithmic divergences do not
occur, and thus this noncommutative supersymmetric gauge theory
reduces to its commutative counterpart when $\Lambda<M_{nc}$.

It is interesting to ask about the limiting case with zero or small
B field. We start from the field theory limit, where
$\ap\rightarrow0$, $\theta$ finite, $B\sim{1\over\theta}$, thus
small values of B correspond to large noncommutative parameter
$\theta$. When gravitational effects are still weak and $\ap$ small
(compared to $\theta$), small values of B correspond to large values
of $\theta$. Our constraint for the validity of the noncommutative
field theory doesn't contradict the existence of vacua with $B=0$ or
very small value of $B$.

We can also consider gauge theory in 2+1 dimensions. It is
interesting at with noncommutativity, even pure $U(1)$ gauge theory
admits finite energy solitonic solutions \dn. The energy is
\eqn\vortex{E={{2\pi}\over{g_3^2}}\int dt \Tr \half F^2, } where
$g_3$ is the U(1) gauge coupling in three dimensions. With vortex
number $n=\Tr F^2$, the simplest nonsingular fluxon solutions have
energy \eqn\fluxon{M_f={{\pi n}\over{g_3^2\theta}}.} When gravity is
turned on, we similarly get a constraint for the gauge coupling
\eqn\vortcon{{{g_3^2\theta}\over n}\geq G_3, } here $G_3$ is the
Newton coupling constant in three dimensions. In the $n=1$ sector,
$M_{nc}=\theta^{-1/2} \leq g_3/\sqrt{G_3}$. Similarly we conjecture
that $\Lambda \leq g_3/\sqrt{G_3}$ for the commutative gauge theory.
Instead of the dimensional gauge coupling $g_3$ we use the
dimensionless gauge coupling ${\tilde g_3}=\Lambda^{-1/2} g_3$
\foot{In \bjs, the dimensionless gauge coupling is defined as
${\tilde g_3}'=M_3^{-1/2}g_3$ and thus $\Lambda\leq {\tilde
g_3}'M_3$. Since $M_3$ is not a reasonable scale if ${\tilde
g_3}<1$, we don't use the definition in \bjs. } and thus
\eqn\tlg{\Lambda\leq {\tilde g_3}^2/G_3\sim {\tilde g_3}^2 M_3.}
Naturalness says that the dimensionless coupling is roughly ${\cal
O}(1)$ and thus $\Lambda \leq M_3$. If there is a UV theory
including gravity beyond the this effective U(1) theory, the value
of ${\tilde g_3}$ is decided by this UV theory and the matching
condition should satisfies $\Lambda\sim {\tilde g_3}^2 M_3$. A
similar argument is also reliable for the result in \amnv.

Next, we consider the noncommutative scalar field theory with
polynomial potential in $\phi$. According to Derrick's theorem \dk,
there is no finite energy solitonic solutions for commutative scalar
field theories in three and more dimensions. But noncommutativity
provides a natural mechanism for stabilizing objects of size
$\sqrt{\theta}$. For sufficiently large $\theta$, there are
solitonic solutions \refs{\dn,\gms,\jh} in odd dimensional scalar
noncommutative field theories with nice potentials. In the
following, we use the notation of \dn\ and show some concrete
examples.

We consider a scalar field theory with Euclidean action in $2+1$
dimensions \eqn\action{S=\int d^3 x {\sqrt g}\(\half
g^{ij}\p_i\phi\p_j\phi+V(\phi)\).} In the canonically commuting
noncommutative coordinates
\eqn\zzbar{\eqalign{z={{x^1+ix^2}\over{\sqrt \theta}} \cr
\overline{z}={{x^1-ix^2}\over{\sqrt \theta}}},} the energy is
\eqn\energy{E=\int d^2 z \(\half(\p\phi)^2+\theta V(\phi) \).} When
$\theta V$ is large, the potential energy dominates and we can find
an approximate solitonic solution by solving the equation ${dV\over
d\phi}=0$.

For an illustrating example, consider a cubic potential
\eqn\vvv{V(\phi)=\half m^2\phi^2+{1\over3}\lambda_3 \phi^3. }
Solving ${dV\over d\phi}=0$ yields $\phi=0, -m^2/\lambda_3$. With
large enough $\theta$ ($\theta
> 1/m^2$), the simplest solitonic solution has energy \refs{\dn,\jh}
\eqn\ener{E_0=2\pi \theta V\(-m^2/\lambda_3\)={{\pi
m^6\theta}\over{3{\lambda_3}^2}}.} We can similarly turn on gravity,
and this soliton will create a deficit angle in spacetime, which
should not exceed $2\pi$ \eqn\deficb{8\pi G_3 E_0\leq 2\pi.}
Neglecting order 1 coefficients in the subsequent analysis, we
get\eqn\gscalar{{{\lambda_3}^2 \over m^6}\geq G_3 \theta.} The cubic
coupling should also be nonzero when space-space is sufficiently
noncommutative. Note that $\lambda_3$ has dimension of length square
and its lower bound is determined by both gravitational effect and
noncommutative effect. In the absence of gravity ($G_3\rightarrow
0$), the bound is trivial. Taking into account the condition for the
existence of the noncommutative soliton $\theta>1/m^2 $ yields
\eqn\mgt{{1\over \theta}\leq m^2 \leq {\lambda_3 \over \sqrt{G_3}}}
in the noncommutative field theory. The right part of eq. \mgt\ is
independent on the noncommutative parameter. The dimensional
quantities depend on the scale $\Lambda$. Define the dimensionless
variables ${\tilde m}=\Lambda^{-1}m$ and ${\tilde
\lambda_3}={\Lambda}^{-3/2}\lambda_3$. Thus eq. \mgt\ becomes
\eqn\cmgt{\Lambda\leq \({\tilde \lambda_3} \over {\tilde m}^2 \)^2
M_3. } For $\Lambda\geq m$, ${\tilde m}\leq 1$; otherwise the
lightest quanta cannot be excited. Naturalness implies ${\tilde
\lambda_3}\sim {\cal O}(1)$ and eq. \cmgt\ becomes $\Lambda \leq
M_3$, which can be accepted by any effective field theorist. The
bound on the scale in eq. \cmgt\ becomes significant only for the
weak coupled situation.

The authors in \lssw\ cited our results to support their
conjectures. In fact, this mechanism is very generic. One can
consider for example another potential of the form
\eqn\quar{V=-{\lambda_6\over 6}\phi^6+{\lambda_8\over 8}\phi^8+{1
\over 24}{\lambda_6^4\over \lambda_8^3},} where we add the last term
in order that the potential at the global minimum equals zero. For
the potential \quar, the global minima are located at $\phi=\pm
\sqrt{\lambda_6/ \lambda_8}$ and the effective mass for the field
theory about one of the minima is
$m_{eff}=\sqrt{2\lambda_6^3/\lambda_8^2}$. When $\theta
m_{eff}^2>1$, there is a noncommutative soliton with energy
\eqn\quare{E\sim {\theta \lambda_6^4\over \lambda_8^3},} and
subsequently the constraint reads \eqn\qqu{{\lambda_8^3\over
\lambda_6^4}\geq G\theta.} Combining the condition for the existence
of the noncommutative soliton yields \eqn\qquu{{1 \over \theta}\leq
{\lambda_6^3 \over \lambda_8^2}\leq {\lambda_6^2\over \lambda_8}
M_3. } The first term in the potential \quar\ is marginal. Define
the dimensionless variable ${\tilde \lambda_8}=\Lambda \lambda_8$.
The right part of eq. \qquu\ becomes \eqn\cqqu{{\lambda_6 \over
\lambda_8}\leq M_3, \quad \hbox{or}, \quad \Lambda \leq {{\tilde
\lambda_8} \over \lambda_6}M_3. } For a general potential, the above
calculations can be easily generalized to give a constraint on the
couplings involved if there is a solitonic solution.

Because of the absence of a principle to constraint the shape of the
potential for the scalar field, we investigate the weak gravity
constraint on the scalar field theory case by case. Here we also
need to remind the reader that the noncommutative scalar field
doesn't simply reduce to commutative scalar field theory for low
momentum \mrs, because of the UV/IR mixture. Whether such effects
will destroy the constraints on general scalar field theory
certainly needs more study, even though the noncommutative parameter
$\theta$ doesn't appear in eq. \cmgt\ and \cqqu. On the other hand,
the conjecture in \lssw\ seems too strong. For example, there is no
 solitonic solution for the 2+1 dimensional scalar field theory with potential
$V=\half m^2\phi^2+{\lambda_4 \over 4}\phi^4$ and with $\lambda_4>0$
if $m^2>0$, even in noncommutative space. There is no evidence to
support the conjecture with $\lambda_4/m^2 \geq G_N$ in \lssw. But
for $m^2=-\mu^2<0$, the potential is just the potential of Higgs
field in the standard model where $\mu/\sqrt{\lambda_4}$ is just the
electroweak energy scale. Requiring that the electroweak energy
scale should be lower than the Planck scale $1/\sqrt{G_4}$ leads to
$\lambda_4/\mu^2\geq G_4$, which is a trivial result.

In this short note we proposed some conjectures on the
noncommutative gauge theory and scalar field theory and related
results for their commutative counterparts. The constraints on the
noncommutative scalar field is only in odd dimensional spacetime.
However we need to keep in mind that gravity does not contain
propagating degrees of freedom in three dimensions. We have no
evidence to support these constraints on the scalar field theories
in four dimensions. On the other hand, whether the noncommutative
solitonic solutions exist when the deficit angle approaches to
$2\pi$ in three dimensions is still an open question. But we believe
that the magnetic monopole in U(1) gauge theory should collapse to
be a black hole when the Newton coupling constant in four dimensions
is large enough because of the conservation of the energy and the
magnetic charge. However there is not a corresponding charge for the
scalar field theory. The constraints on the scalar field theories
need more study in the future. A more ambitious plan is to
investigate the implication on the inflaton potential from the
viewpoint of weak gravity conjecture on the scalar field theory.

\bigskip

Acknowledgments.

We would like to thank M. Li, F.L. Lin, W. Song, J.B. Wu and X.H. Wu
for useful discussions.

\listrefs
\end